\documentclass[fleqn,12pt,twoside]{article}
\usepackage[headings]{espcrc1}

\readRCS
$Id: espcrc1.tex,v 1.2 2004/02/24 11:22:11 spepping Exp $
\usepackage{graphicx}
\usepackage[figuresright]{rotating}

\newcommand{\AmS}{{\protect\the\textfont2
  A\kern-.1667em\lower.5ex\hbox{M}\kern-.125emS}}

\title{
Subbarrier fusion reactions with dissipative couplings}

\author{K. Hagino
\address[a]{Department of Physics, Tohoku University, 
Sendai 980-8578, Japan}, 
S. Yusa
\addressmark[a],
and N. Rowley
\address{Institut de Physique Nucl\'eaire, 91406 Orsay Cedex, France}}
       
\runtitle{Subbarrier fusion reactions with dissipative couplings}
\runauthor{K. Hagino, S. Yusa, and N. Rowley}

\begin{document}

% typeset front matter
\maketitle

\begin{abstract}
Using the random matrix model, we discuss 
the effect of couplings to non-collective states on the penetrability 
of a one dimensional potential barrier. 
We show that these non-collective excitations hinder the penetrability 
and thus smear the barrier distribution at 
energies above the barrier, while they do not affect significantly 
the penetrability at 
deep subbarrier energies. The energy dependence of the 
$Q$-value distribution obtained with this model is also discussed. 
\end{abstract}

\section{Introduction}

The coupled-channels approach 
has been successful in describing the subbarrier enhancement of 
heavy-ion fusion
cross sections ~\cite{DHRS98,BT98}. 
Conventionally, it takes into account the coupling between the relative motion 
of the colliding nuclei and a few low-lying 
collective excitations in the colliding 
nuclei, as well as transfer channels, 
which couple strongly to the ground state. 
High-lying modes, such as giant resonances, and single-particle excitations 
are not considered usually, since the former simply renormalizes the 
internucleus potential ~\cite{THAB94} 
and the latter is not coupled strongly to the 
ground state. 

However, in recent years,
many 
experimental data have accumulated that suggest a need to go beyond the
conventional coupled-channels approach. The examples include 
the surface diffuseness anomaly in the internuclear potential~\cite{N04}, 
the steep fall-off
of fusion cross sections at deep subbarrier energies~\cite{J02,D07,S08}, a
large smoothing of quasi-elastic barrier distribution~\cite{T95,P05}, and
the energy dependence of the $Q$-value spectra for quasi-elastic back
scattering~\cite{E08,L09}.
It has been a challenge to account for these new aspects of heavy-ion 
fusion reactions simultaneously with the coupled-channels framework. 

Recently, quasi-elastic scattering cross sections 
for $^{20}$Ne+$^{90,92}$Zr systems at backward angles have been measured, which show a considerable difference 
in the barrier distribution between the two systems ~\cite{E09}, that is, 
the barrier distribution with the $^{92}$Zr target is much more smeared than that with $^{90}$Zr. The 
coupled-channels calculations, on the other hand, predict a similar barrier distribution 
to each other for both the systems, 
because the rotational excitations of $^{20}$Ne play a 
predominant role. Since those coupled-channels calculations include 
the collective excitations in the $^{90,92}$Zr nuclei, the experimental 
data strongly indicate that the difference in the barrier distribution for the two 
systems can be attributed to non-collective excitations in the target nuclei. 
Notice that 
the effect of single-particle excitation should be more important in $^{92}$Zr, 
compared to a $N=50$ magic nucleus, $^{90}$Zr. 

In this
contribution, we shall 
discuss the effect of single-particle excitations on heavy-ion reactions, that
have been ignored in the conventional coupled-channels approach. 
To this end, we compute 
the penetrability for a one dimensional
two-level system in the presence of a coupling to dissipative environment
described by a random matrix model. 
The random matrix model was originally developed by Weidenm\"uller and his
collaborators in the late 70's in order to describe deep inelastic
collisions for massive systems~\cite{AKW77}. 
Here we shall use a similar model,
and solve quantum mechanically
the coupled-channels equations of a large dimension.
See Refs. ~\cite{HT98,DT08,Z90} for 
earlier attempts, which however did not use the random matrix model. 

\section{One-dimensional barrier penetrability with random matrix model}

In the random matrix model, one considers an ensemble of 
the coupling matrix elements, $V_{ij}(x)$, in the coupled-channels 
equations,  
which are assumed to follow the Gaussian Orthogonal Ensemble (GOE) 
~\cite{AKW77}. 
That is, they have a zero mean, $\overline{V_{ij}(x)}=0$, 
and the second moment is 
given by 
\begin{equation}
\overline{V_{ij}(x)V_{kl}(x')}=
(\delta_{i,k}\delta_{j,l}+\delta_{i,l}\delta_{j,k})\,
\frac{w_0}{\sqrt{\rho(\epsilon_i)\rho(\epsilon_j)}}\,
e^{-\frac{(\epsilon_i-\epsilon_j)^2}{2\Delta^2}}\,
\cdot e^{-\frac{(x-x')^2}{2\sigma^2}}\cdot e^{-\frac{x^2+x'^2}{2\alpha^2}},
\end{equation}
where $\rho(\epsilon)$ is the level density. 

We apply this model to a one-dimensional system ~\cite{KPW76}, in which 
we consider a Gaussian potential barrier with the height of 100 MeV
~\cite{DLW83,HB04}. 
In order to take into account the quasi-continuum single-particle spectrum, 
we descretize it ~\cite{N79} from 3 MeV to 13 MeV with an energy spacing 
of 0.05 MeV (in this way, we include 200 single-particle channels). 
In order to take an ensemble average, we generate 20 random matrices and 
perform coupled-channels calculations 20 times for each energy. 
In addition to the single-particle levels, we consider also a 
collective level at 1 MeV, whose coupling form factor is given 
by a Gaussian function ~\cite{DLW83,HB04}. 
The coupling strength to the collective state is set to be the same 
for all the samples in the random ensemble. 

\begin{figure}[htb]
\begin{minipage}[t]{80mm}
\includegraphics*[width=15pc]{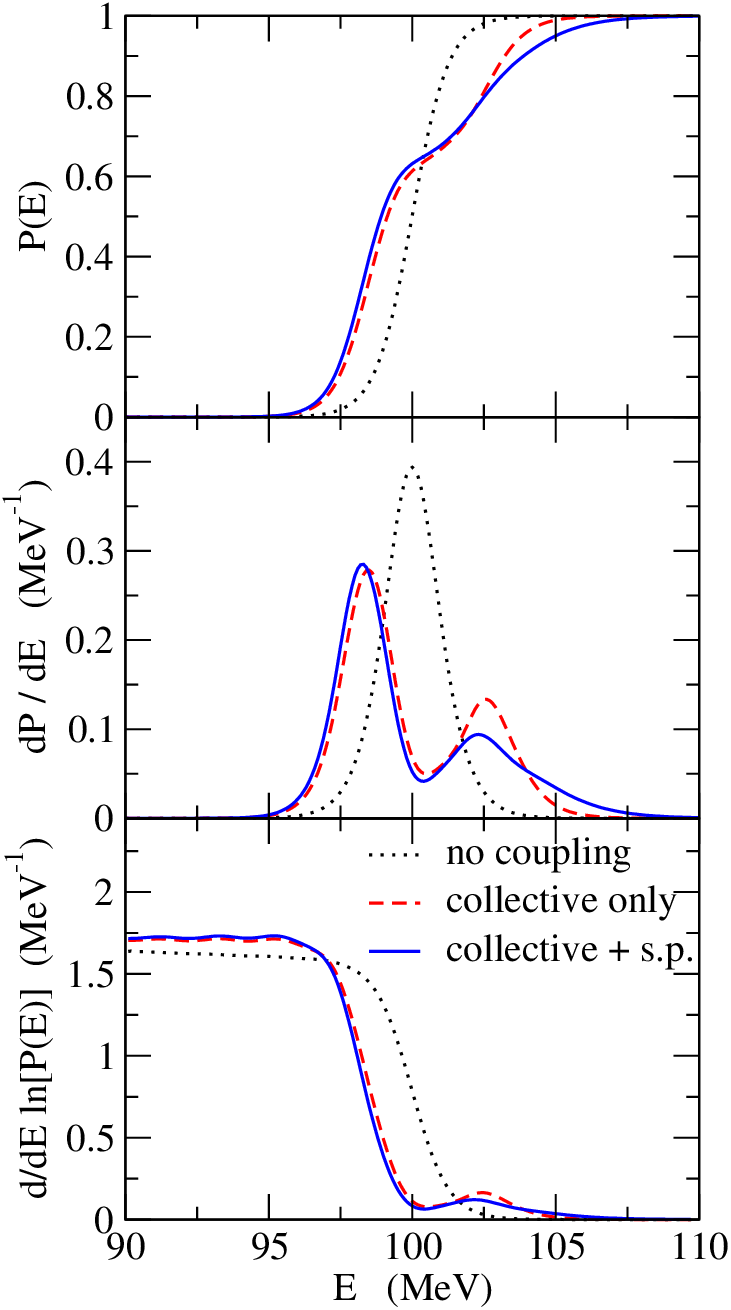}
\caption{The penetrability (the top panel), the barrier distribution 
(the middle panel), and the logarithmic slope of the penetrability (the 
bottom panel) obtained with the random matrix model. }
\end{minipage}
\hspace{\fill}
\begin{minipage}[t]{75mm}
\includegraphics*[width=15pc]{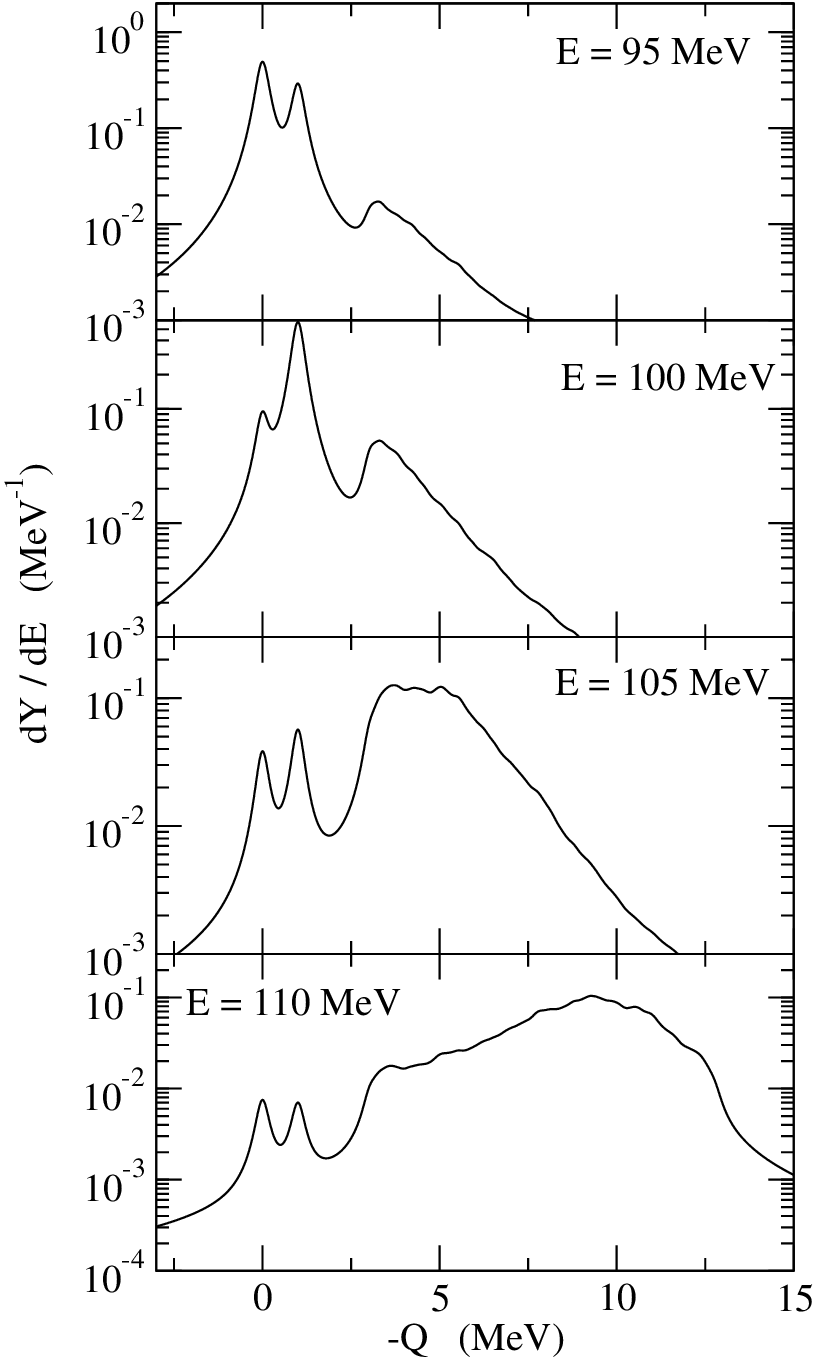}
\caption{The energy dependence of the $Q$-value distribution 
defined with the reflected flux. The peaks at $Q$=0 and $-$1 MeV correspond 
to the elastic scattering and the excitation of the collective state, 
respectively. }
\end{minipage}
\end{figure}

The top panel of Fig. 1 shows the penetrability thus obtained (the solid line), in comparison to 
that without the couplings to non-collective states (the dashed line). The figure also shows 
by the dotted line the result of a single-channel calculation. The effect of 
the non-collective couplings mainly appears at energies above the barrier, where 
the couplings hinder the penetrability. The middle panel shows the barrier distribution ~\cite{DHRS98,BT98}, 
defined as the first derivative of the penetrability. 
In the absence of the non-collective couplings, the barrier distribution has two peaks, corresponding to 
the two eigen-barriers generated from superpositions of the ground and the collective states. 
When the non-collective couplings are switched on, the higher peak is smeared significantly, although the structure 
of the lower peak remains almost the same. 
The bottom panel shows the logarithmic slope of the penetrability ~\cite{J02}, which provides a useful means to 
investigate the deep subbarrier behavior of the penetrability. 
One can see that the logarithmic slope is not altered much by the non-collective couplings, indicating that  
the steep fall-off phenomena of deep subbarrier fusion cross sections do not seem to be accounted for by the 
present mechanism (see also Ref. ~\cite{R09}). 

Figure 2 shows the $Q$-value distribution obtained with the reflected flux 
in the solution of coupled-channels equations. 
We have smeared the discrete distribution with a Lorenzian function with the width of 0.2 MeV. 
At energies below the barrier, only the elastic and the collective channels are important. 
As energy increases, one can clearly see that the single-particle excitations gradually become important, in 
accordance with the results shown in Fig. 1. 
One can also define the $Q$-value distribution with the transmitted flux (not shown). Our calculation indicates 
that the $Q$-value distribution obtained with the transmitted flux is much less sensitive to the non-collective 
couplings compared with the $Q$-value distribution defined with the reflected flux, suggesting 
that quasi-elastic scattering is more sensitive to the single-particle excitations than fusion. 

\section{Summary}

We have applied the random matrix model to a one-dimensional barrier penetrability in order to discuss 
the effect of single-particle excitations. 
We have shown that the effects of non-collective excitations mainly affect the above barrier behavior 
of the penetrability, that is, they hinder the penetrability and smear the barrier distribution. On the 
other hand, the low energy behavior does not appear significant. 
The coupled-channels approach with the random matrix model enables one to compute the $Q$-value distribution. 
We have shown that the single-particle excitation gradually becomes important as energy increases. 

In this contribution, we have used a schematic one-dimensional model. It will be an interesting future 
project to apply this model to realistic systems, and investigate the effect of non-collective excitations 
on quasi-elastic barrier distributions. A quantum mechanical description of deep inelastic collisions 
using the present model will also be 
of interest. 

\section*{Acknowledgement}
We thank E. Piasecki and V.I. Zagrebaev for useful discussions. 
This work was supported by the Japanese
Ministry of Education, Culture, Sports, Science and Technology
by Grant-in-Aid for Scientific Research under
the program number 19740115.

\end{document}